\documentclass[final,3p,times,twocolumn]{elsarticle}

\usepackage[utf8]{inputenc}
\usepackage{amssymb}
\usepackage{graphicx}
\usepackage{hyperref}
\usepackage{siunitx}

\graphicspath{{figures/}}

\begin{document}
\begin{frontmatter}
\title{Performance of CMOS pixel sensor prototypes in ams H35 and aH18
  technology for the ATLAS ITk upgrade\tnoteref{license}}

\author[geneva]{Moritz Kiehn\corref{corresponding}}
\ead{moritz.kiehn@cern.ch}
\author[geneva]{Francesco Armando Di Bello}
\author[geneva]{Mathieu Benoit}
\author[barcelona]{Raimon Casanova Mohr}
\author[bnl]{Hucheng Chen}
\author[bnl]{Kai Chen}
\author[geneva]{Sultan D.M.S.}
\author[kit]{Felix Ehrler}
\author[geneva]{Didier Ferrere}
\author[oklahoma]{Dylan Frizell}
\author[geneva]{Sergio Gonzalez Sevilla}
\author[geneva]{Giuseppe Iacobucci}
\author[bnl]{Francesco Lanni}
\author[bnl,hefei]{Hongbin Liu}
\author[bern]{Claudia Merlassino}
\author[anl]{Jessica Metcalfe}
\author[bern]{Antonio Miucci}
\author[kit]{Ivan Peric}
\author[kit]{Mridula Prathapan}
\author[kit]{Rudolf Schimassek}
\author[geneva]{Mateus Vicente Barreto}
\author[geneva]{Thomas Weston}
\author[liverpool]{Eva Vilella Figueras}
\author[kit]{Alena Weber}
\author[bern]{Michele Weber}
\author[geneva]{Winnie Wong}
\author[bnl]{Weihao Wu}
\author[geneva]{Ettore Zaffaroni}
\author[kit]{Hui Zhang}
\author[urbana]{Matt Zhang}

\address[geneva]{
  Département de Physique Nucléaire et Corpusculaire,
  Université de Genève, 24 quai Ernest Ansermet, 1211~Genève~4, Switzerland}
\address[anl]{
  Argonne National Laboratory, Argonne, IL 60439, USA}
\address[bnl]{
  Brookhaven National Laboratory, P.O. Box 5000, Upton, NY 11973-5000, USA}
\address[barcelona]{
  Institut de Física d’Altes Energies,
  The Barcelona Institute of Science and Technology,
  Edifici CN, UAB campus,
  08193 Bellaterra (Barcelona), Spain}
\address[bern]{%
  Albert Einstein Center for Fundamental Physics and Laboratory for High Energy Physics,
  University of Bern,
  Siedlerstrasse~5, CH-3012 Bern, Switzerland}
\address[hefei]{
  Department of Modern Physics,
  University of Science and Technology of China, Hefei, Anhui 230026, China}
\address[liverpool]{
  Department of Physics,
  University of Liverpool,
  The Oliver Lodge Laboratory,
  Liverpool L69 7ZE, UK}
\address[kit]{
  Karlsruhe Institute of Technology, IPE, 76021 Karlsruhe, Germany}
\address[oklahoma]{
  University of Oklahoma, 660 Parrington Oval, Norman, OK 73019, USA}
\address[urbana]{
  University of Illinois Urbana Champaign,
  1110 W Green St Loomis Laboratory,
  Urbana, IL 61801, USA}

\tnotetext[license]{© 2018. This manuscript version is made available
  under the \href{http://creativecommons.org/licenses/by-nc-nd/4.0/}{CC-BY-NC-ND 4.0} license.}
\cortext[corresponding]{Corresponding author}

\begin{abstract}
  Pixel sensors based on commercial high-voltage CMOS processes are an
  exciting technology that is considered as an option for the outer
  layer of the ATLAS inner tracker upgrade at the High Luminosity LHC.
  Here, charged particles are detected using deep n-wells
  as sensor diodes with the depleted region extending into the silicon
  bulk. Both analog and digital readout electronics can be added to
  achieve different levels of integration up to a fully monolithic
  sensor. Small scale prototypes using the ams CMOS technology have
  previously demonstrated that it can achieve the required radiation
  tolerance of \SI{e15}{n_\text{eq}\per\cm\squared} and detection
  efficiencies above \SI{99.5}{\percent}. Recently, large area
  prototypes, comparable in size to a full sensor, have been produced
  that include most features required towards a final design: the
  H35demo prototype produced in ams H35 technology that supports both
  external and integrated readout and the monolithic ATLASPix1
  pre-production design produced in ams aH18 technology. Both chips are
  based on large fill-factor pixel designs, but differ in readout
  structure. Performance results for H35DEMO with capacitively-coupled
  external readout and first results for the monolithic ATLASPix1 are
  shown.
\end{abstract}
\end{frontmatter}

\section{Introduction}

Tracking detectors at future colliders have to fulfil increasingly
demanding requirements. Their environment will contain a higher density
of tracks, at higher rate, with a higher radiation dose over their
lifetime compared to current detectors. This necessitates high
granularity detectors even far away from the interaction point and
consequently large instrumented surface areas. The planned ATLAS Inner
Tracker (ITk) upgrade \cite{ATLAS-TDR-025,ATL-COM-ITK-2017-073} is one such
detector that is designed to withstand the environment at the planned
high luminosity large hadron collider. Sensors for the outer layer of
the ATLAS ITk pixel tracker have to cope with an integrated radiation
dose of up to \SI{e15}{n_\text{eq}\per\cm\squared} over its lifetime.

Silicon pixel detectors are the only technology that can provide the
granularity, rate capability, and radiation hardness. Traditionally they
are implemented as hybrid pixel sensors. Traversing
particles generate free charges in a fully depleted passive sensor
diode. Pixelated electrodes on the sensor are connected via bump-bonding
to a dedicated readout chip. The readout chip contains both analog
amplifiers and digital processing logic and handles the triggering and
readout.

While this separation of concerns allows each component to be optimized
separately it also introduces additional challenges. A large
charge signal is required to generate a large enough voltage signal over
the total capacitance of the sensor diode and the readout electronics.
This necessitates a thick, fully depleted sensor diode with a
high voltage bias to deplete it. As an example, the central modules used
in the ATLAS IBL detector are built from \SI{200}{\um} thick planar
sensors, with bias voltages foreseen to reach \SI{1}{\kV}, bump-bonded
to a \SI{150}{\um} thick readout chips \cite{ATLAS-TDR-19}.
The complexity of the production process,
i.e. separate sensor and readout production and hybridization, can be a
limiting factor to production capabilities. Yield factors compound and
due to the many steps involved the production can not easily scale to
large number of sensors and large instrumented surface areas.

Pixel sensors based on CMOS technology enable simpler devices with a
reduced material budget by integrating some or all of the readout
electronics directly into one chip. Commercial production enables cheap
sensors with greatly simplified production complexity, suitable to
instrument large surface areas with high granularity sensors.

In the remainder of this paper, basic concepts are introduced and
different implementations of CMOS-based pixel sensors using ams
technology are presented. Then, specific prototypes and the status of
ongoing prototype evaluations are discussed.

\section{ams CMOS prototypes}

The ams technology is a set of commercial CMOS processes that are
available in \SI{350}{\nm} and \SI{180}{\nm} structure sizes
\cite{amshvcmos}. All processes can use nMOS and pMOS transistors and
support high voltages of up to \SI{120}{\V}. High-resistivity substrates
up to \SI{1000}{\ohm\cm} are also available.

Previous small-scale prototypes based on this technology used integrated
amplifiers and comparators in combination with an external readout
chip. They could be operated with efficiencies above \SI{99.5}{\percent}
for fluences up to \SI{1.5e15}{n_\text{eq}\per\cm\squared} using both
neutron and proton irradiation
\cite{ccpdv4}, demonstrating the radiation hardness of this
technology. Other small-scale prototypes demonstrated the feasibility of
integrating different levels of readout logic onto the same chip
\cite{Augustin:2015mqa,Augustin:2016hzx}.

\begin{figure}
  \includegraphics[width=\linewidth]{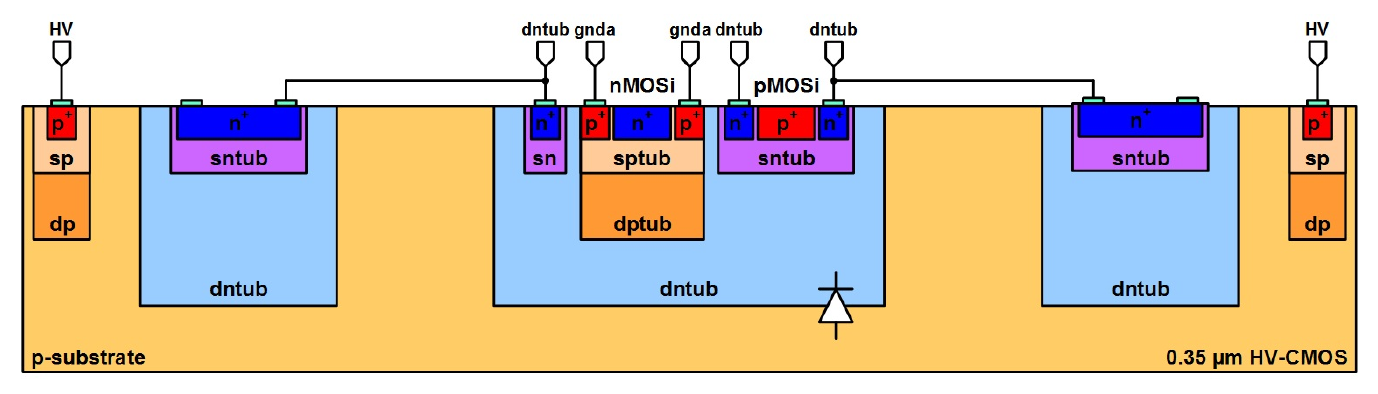}
  \caption{Cross-section of the pixel implants for the H35DEMO prototype
    along the large \SI{250}{\um} pitch direction. Three separate sensor
    diodes (deep n-well labeled as dntub) are connected together to
    reduce the total sensor capacitance while maintaining a homogeneous
    depletion. The additional deep p-type implant (labeled as dptub)
    underneath the nMOS components is optional and only present in some
    parts of the test matrices.}
  \label{fig:h35demo_implants}
\end{figure}

\begin{figure}
  \includegraphics[width=\linewidth]{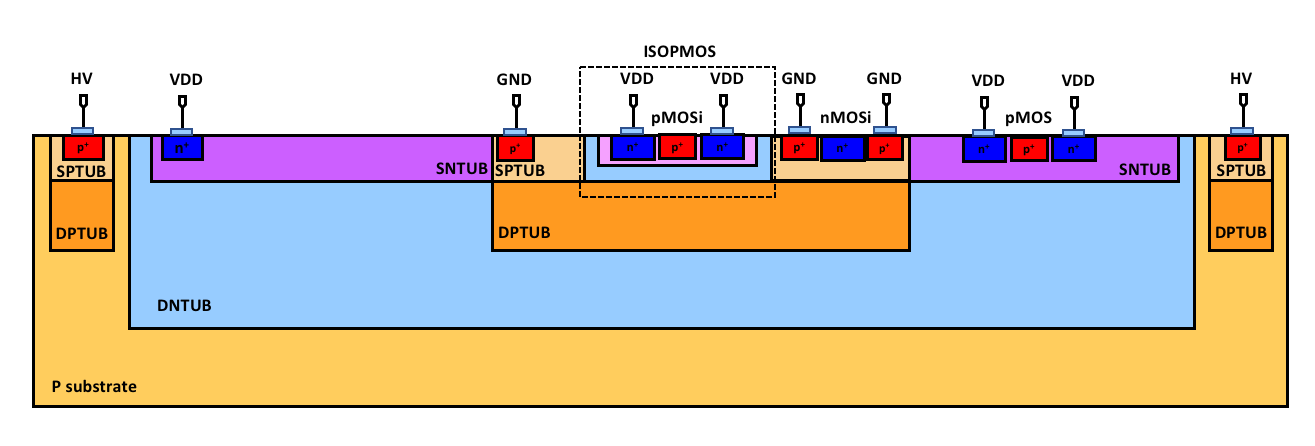}
  \caption{Cross-section of the pixel implants for ATLASPix1 prototype
    along the large \SI{140}{\um} pitch direction. The additional deep
    p-type implant (labeled as dptub) is only present in the IsoSimple
    matrix.}
  \label{fig:atlaspix1_implants}
\end{figure}

Two large-scale pixel sensor prototypes, H35DEMO and ATLASPix1, have
been recently produced to show the viability of this technology for
full-sized production sensors as an upgrade option for ATLAS ITk and
to test different aspects of the technology.
All prototypes are based on so-called large fill-factor designs. The
large fill-factor refers to the size of the sensor diode compared
to the pixel pitch. In these designs, the pixel electronics are located
inside a deep n-type implant in a p-type substrate. The deep n-type
implant and the substrate form the sensor diode that is depleted by
applying a bias voltage. Usually bias from the top side is used;
back-bias could also be employed but requires additional back-side
processing. Figure \ref{fig:h35demo_implants} shows a cross-section of
the pixel implants for the H35DEMO prototype along the large
pitch direction. Here, three separate sensor diodes are connected
together to reduce the total sensor capacitance while maintaining a
homogeneous depletion. Figure \ref{fig:atlaspix1_implants} shows the
equivalent pixel implants cross-section for the ATLASPix1 prototype. Due
to a smaller size only a single diode implantation is used.

\begin{figure}
  \includegraphics[width=\linewidth]{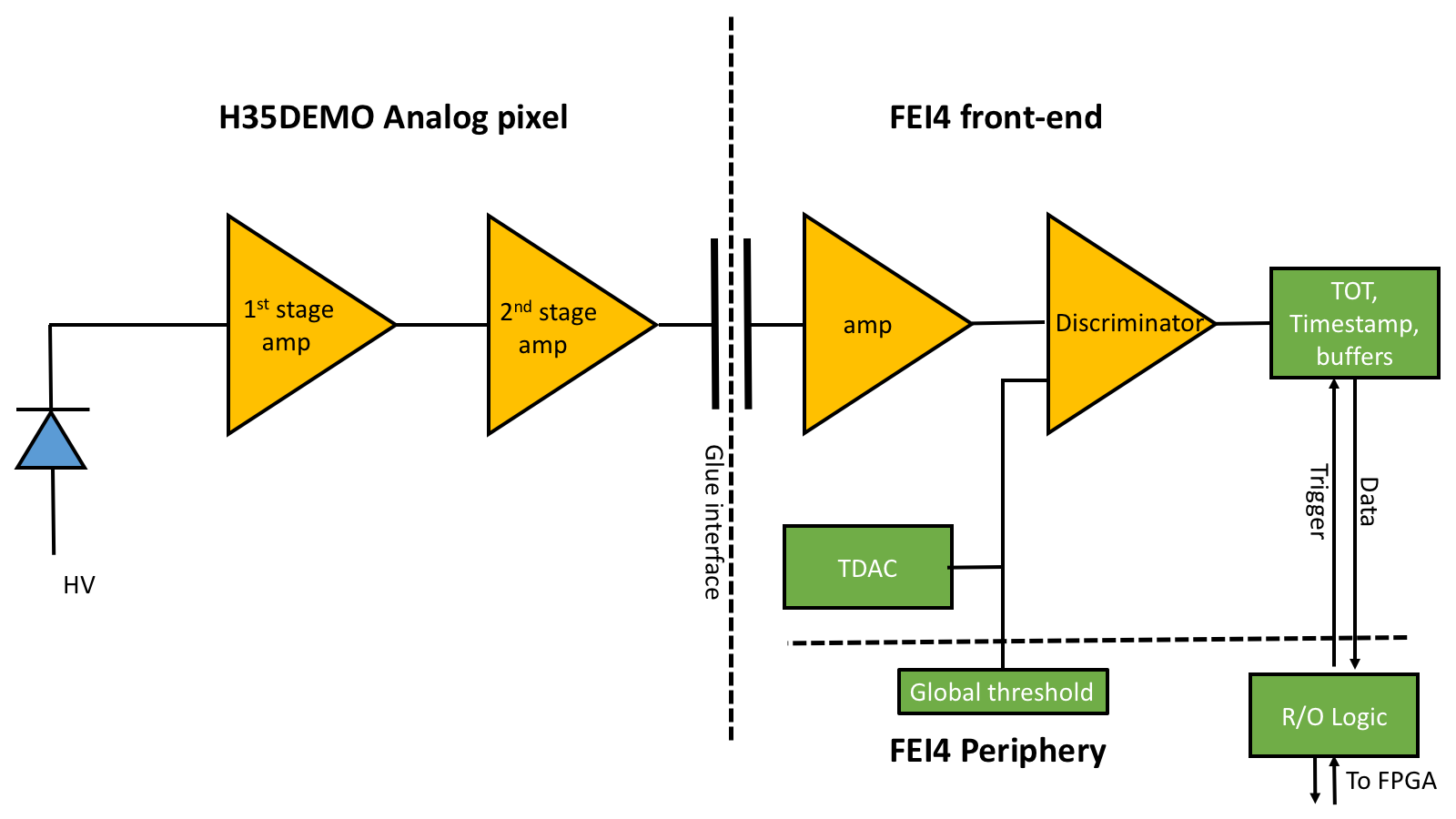}
  \caption{The signal path on the H35demo prototype. The two amplifier
    stages before the glue interface are implemented in ams H35 CMOS
    technology on the prototype chip. The glue interface capacitively
    couples the amplified signal to the FE-I4 readout chip.}
  \label{fig:h35demo_signal_path}
\end{figure}

H35DEMO is produced in H35 \SI{350}{\nm} technology with a total size of
\SI{24.4x18.5}{\mm} and a pixel pitch of \SI{250x50}{\um} \cite{h35demo:design}. It
comprises four independent matrices: two monolithic matrices with
integrated readout and two analog matrices that integrate only per-pixel
amplifiers. On the first monolithic matrix each pixel contains
amplifiers and comparator while on the second matrix only the amplifiers
are located in the pixel and the comparator is located in the chip
periphery.
Evaluation results for the two monolithic matrices were previously
reported by \citet{Cavallaro:2016gmx} and \citet{Terzo:2017hlv}.
The analog matrices provide an amplified analog signal that
is then readout by a capacitively coupled FE-I4 readout chip as shown in
figure \ref{fig:h35demo_signal_path}. The two analog matrices differ in
the layout of their in-pixel electronics. Within each analog matrix
there are different submatrices with additional small variations of the integrated
electronics.

\begin{figure}
  \includegraphics[width=\linewidth]{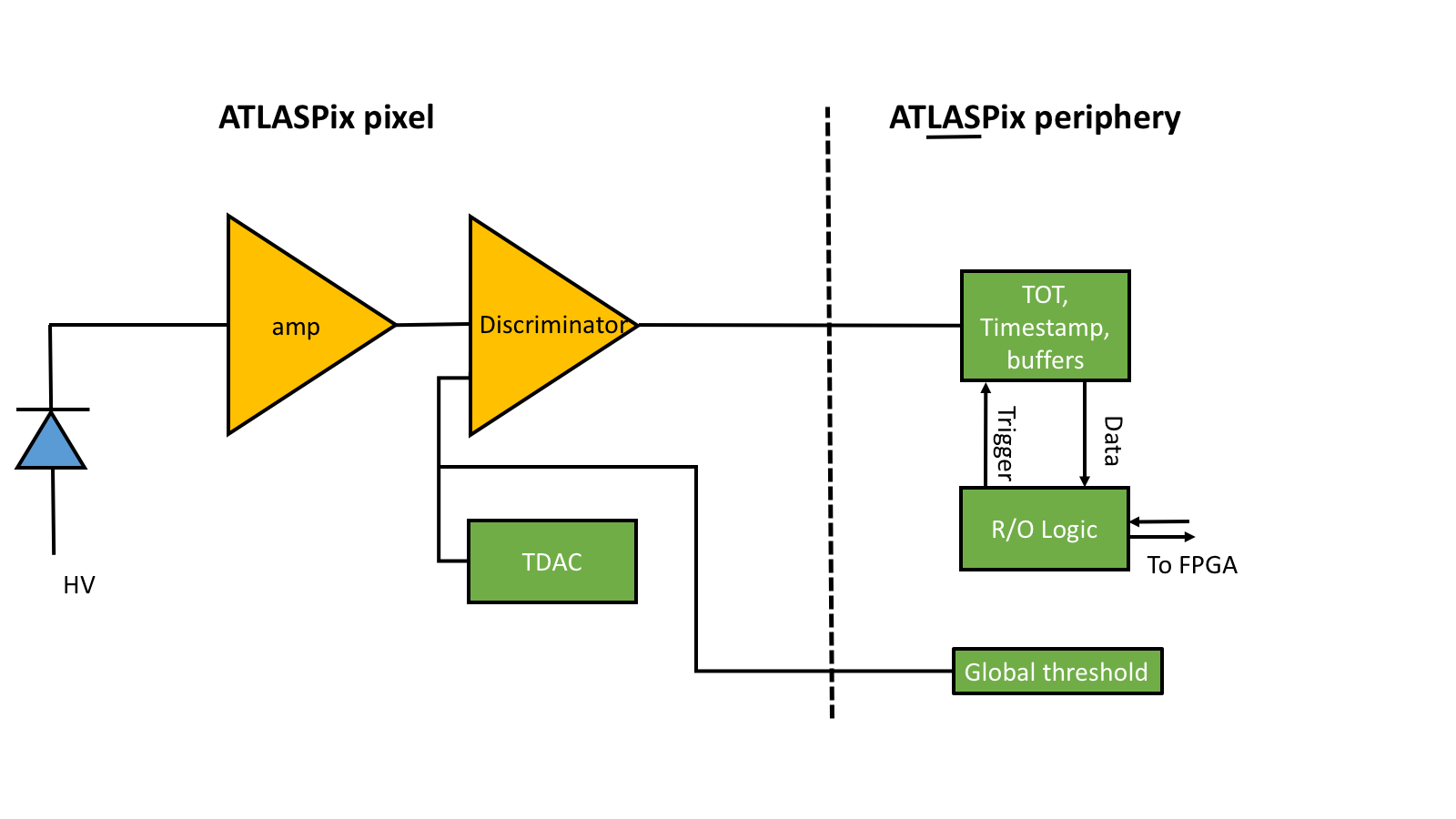}
  \caption{The signal path on the ATLASPix1 prototype. All components
    are implemented in ams aH18 CMOS technology on the prototype
    chip. The amplifier and the discriminator are located inside the
    pixel. Each pixel is connected with dedicated lines to the periphery
    of the chip which contains the remaining readout components.}
  \label{fig:atlaspix1_signal_path}
\end{figure}

ATLASPix1 is a monolithic prototype with integrated readout logic and a
total size of approximately \SI{1x2}{\cm} produced in aH18 \SI{180}{\nm}
technology. It comprises three independent matrices --- M2, Simple and
IsoSimple --- that differ in readout architecture and pixel pitch. M2 has a
pixel pitch of \SI{60x50}{\um} and uses a triggered readout
architecture. Simple and IsoSimple have a pixel pitch of
\SI{140x40}{\um} and use an untriggered column-drain architecture. The
IsoSimple matrix has an additional isolation p-well, as shown in figure
\ref{fig:atlaspix1_implants}, and use full CMOS transistor in its
comparator logic. The Simple matrix only uses nMOS logic and misses the
isolation p-well. As shown in figure \ref{fig:atlaspix1_signal_path},
all matrices have the per-pixel amplifier and comparator located inside
each pixel which are connected to additional logic in the digital
periphery.

\section{Beam tests}

The operational performance of both prototypes was tested in beam test
setups. Measurements with the H35DEMO were performed in spring 2017 at
the Fermilab MTEST facility using the \SI{120}{\GeV} primary proton beam
and in summer 2017 at the CERN north area beam facilities using a
\SI{180}{\GeV} secondary mixed hadron beam. Measurements with the
ATLASPix1 prototype were performed in October 2017 at the CERN north
area facilities also using a \SI{180}{\GeV} secondary mixed hadron beam.

Reference tracks were measured using the Geneva beam telescope. It uses
six ATLAS IBL modules with a pixel pitch of \SI{250x50}{\um} to measure
particle positions \cite{genevatelescope}. The sensors are arranged in
an optimized geometry with rotated and inclined planes to maximize
charge sharing and enhance the resolution. The telescope and FE-I4-based
devices-under-test are readout using the RCE/HSIO2 data acquisition
system \cite{genevatelescope}. Triggers are provided by hit coincidences
on the first and the last plane with a variable integration time between
8 and 16 time bins of \SI{25}{\ns}.

The H35DEMO prototypes with capacitively coupled FE-I4 readout sensors
can be read out directly using the RCE/HSIO2 system. The ATLASPix1 is
operated with an independent data acquisition system based on a
commercial Xilinx Nexys FPGA board in combination with a custom control
and readout board developed at KIT. It is controlled via USB2 using a
custom software. Integration into the telescope system is achieved via
a trigger-busy scheme. The trigger signal from the telescope is used by
the FPGA to write data only for triggered events into a separated
stream. Synchronisation between the two systems happens offline based on
trigger counting and timestamps.

Particle tracks are reconstructed with the Proteus reconstruction
software \cite{proteus} using only the data from the telescope
planes. Proteus performs hit clustering, weighted center-of-gravity
position, and detector alignment. Initial coarse alignment is based on
hit correlations and subsequent fine alignment uses track
residuals. Reconstructed track positions and the full reconstruction
covariance matrix are propagated into the local system of the
device-under-test and are then used to calculated residuals and
efficiency.

\section{H35DEMO results}

A systematic evaluation of the performance of the two H35DEMO analog
matrices has been performed. In this configuration only analog
electronics, i.e. signal amplifiers and shapers, but no digital readout
logic is integrated on the CMOS chip. It can therefore be used to test
the performance of the sensor and the analog electronics
independent of the readout logic by using the well-tested FE-I4 readout
chip. Detailed evaluation results for the H35DEMO analog matrices have
previously been reported by \citet{h35demo:results}. Only a
high-level overview is provided here to give the reader a full picture
of all ams CMOS prototypes.

\begin{figure}
  \centering
  \includegraphics[width=\linewidth]{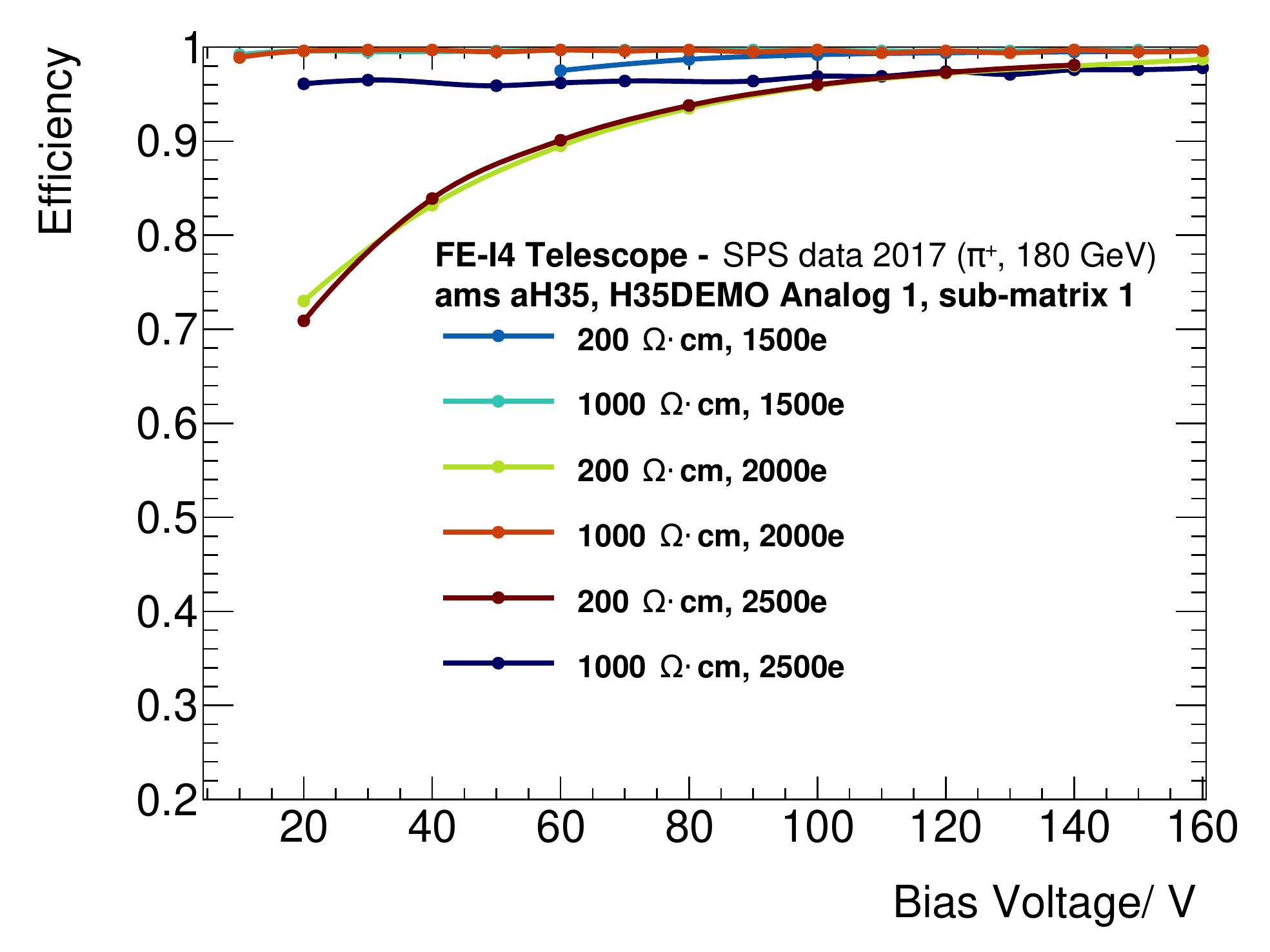}%
  \caption{H35demo efficiency measurements for one sub-matrix of the
    first analog matrix for different resistivities and
    thresholds. \cite{h35demo:results}}
  \label{fig:h35demo_ana1}
\end{figure}

Figure \ref{fig:h35demo_ana1} shows the resulting global efficiency
measured for one submatrix of the first analog matrix for different
device configurations as a function of the sensor bias voltage. The
threshold quoted is the threshold set on the coupled FE-I4 readout chip. Since
the initial signal is already amplified and shaped by the integrated
amplifiers on the H35DEMO the effective threshold in terms of generated
charge in the sensor is smaller. As expected, a clear dependence is seen
on both substrate resistivity and threshold. A higher substrate
resistivity leads to larger signal and therefore to a higher efficiency
for the same threshold and bias. The same is true for lower
threshold. For both shown substrate resistivities an operational region with
efficiencies above \SI{99.5}{\percent} can be found.
The measured efficiency is comparable to or better than results from
traditional hybrid modules with bump-bonded passive sensors
and fulfil the efficiency requirement of \SI{99}{\percent}
for the outer layer of the ATLAS ITk upgrade \cite{ATL-COM-ITK-2017-073}.

\section{ATLASPix1 results}

\begin{figure}
  \centering
  \includegraphics[width=\linewidth]{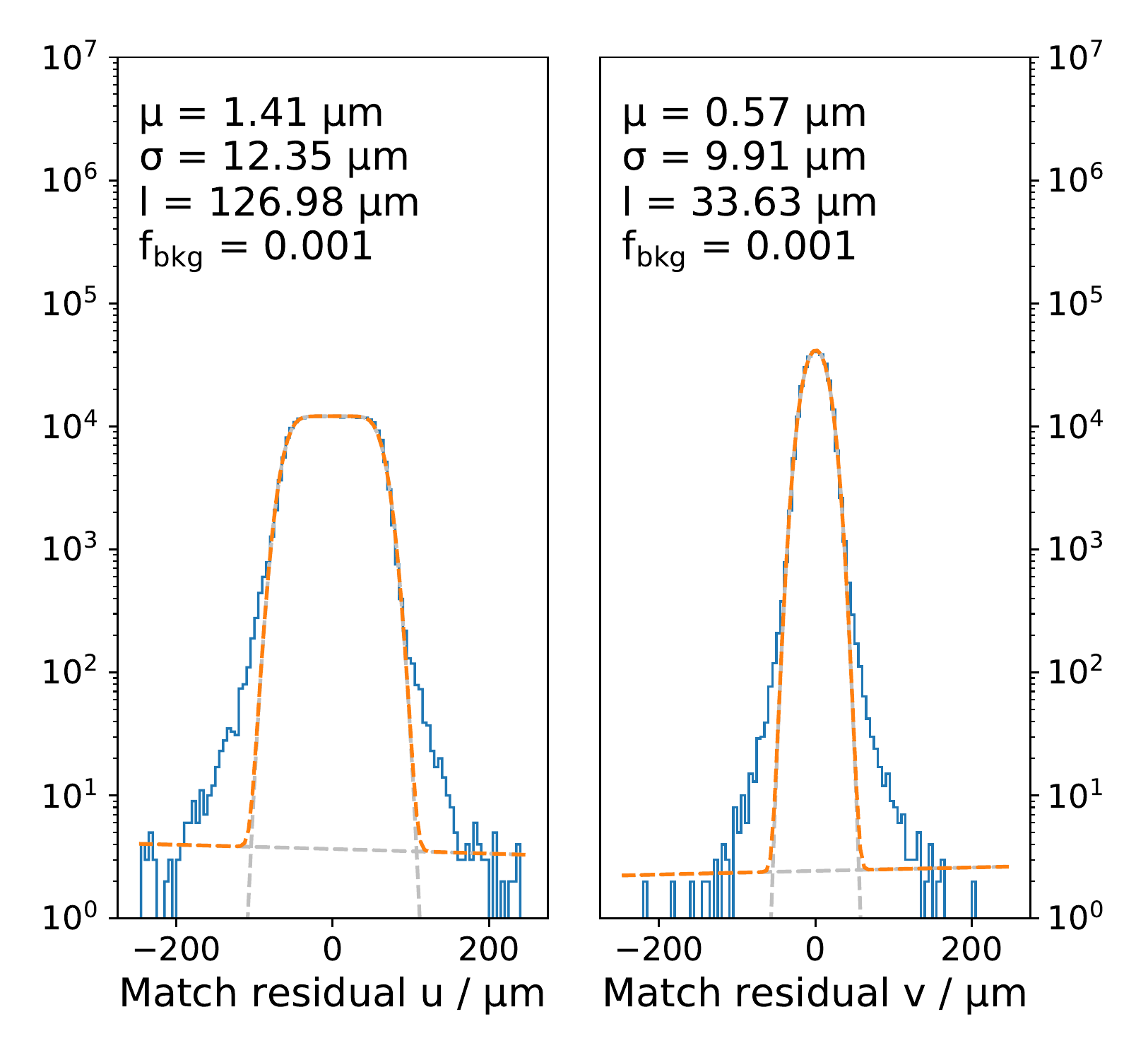}%
  \caption{ATLASPix1 Simple matrix matching residuals for the the long u
    and the short v pixel axis for a sensor bias of \SI{65}{\V} and a
    threshold of \SI{840}{\mV}. Residuals are calculated between the
    extrapolated particle positions on the device-under-test as
    reconstructed by the beam telescope and the center-of-gravity of the
    cluster. The fitted function is a box function of width l smeared
    with a Gaussian of with $\sigma$ and a polynomial background.}
  \label{fig:atlaspix1_residuals}
\end{figure}

\begin{figure}
  \centering
  \includegraphics[width=\linewidth]{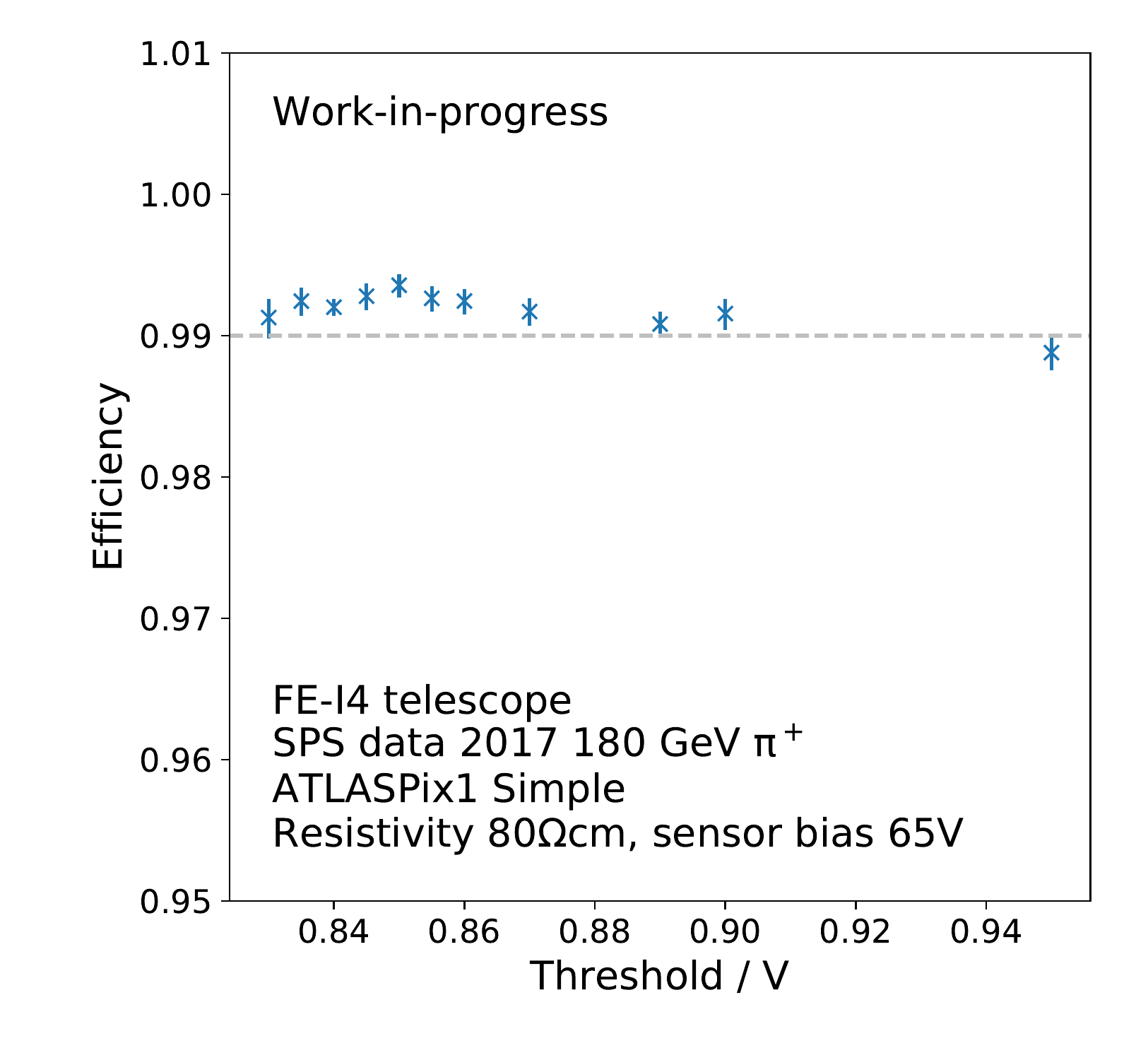}
  \caption{ATLASPix1 simple matrix efficiency scan as a function of the
    global threshold for a fixed sensor bias of \SI{65}{\V}. The
    threshold baseline is \SI{800}{\mV}. The error bars represent only
    the statistical uncertainties. Preliminary calibrations indicate
    that the measured threshold range corresponds to an equivalent
    charge range of approximately \SIrange{300}{1100}{e}.}
  \label{fig:atlaspix1_efficiency}
\end{figure}

Given the availability of prototype samples and time constraints
measurements could only be performed for the ATLASPix1 Simple matrix of
one unirradiated sample with a substrate resistivity of
\SI{80}{\ohm\cm}. These measurements test both the analog performance of
the sensor and the integrated readout logic.

Figure \ref{fig:atlaspix1_residuals} shows the residuals between
reconstructed track position and estimated cluster position on the
device-under-test for one particular configuration. A matching cut of
\SI{250}{\um} along both dimensions is used to associate tracks to
clusters. The fitted function models the expected response from a pixel
of width $l$ assuming a telescope resolution of $\sigma$. The vanishing
mean $\mu$ along both axis shows that the system is well-aligned. The
fitted pixel width along the long u direction is consistent with
the \SI{140}{\um} pitch. Along the short direction v the fitted width
is smaller than the \SI{40}{\um} pitch as a result of increased charge
sharing. In both cases the background fraction $f_{\text{bkg}}$ due to
mismatches or noise hits is negligible.

Using tracks and clusters within the matching cut, the global efficiency
can be calculated as the fraction of tracks with an associated cluster
and the total number of tracks. Tracks are only considered within a
selected region-of-interest in the central part of the sensors. This
avoids edge effects, low statistic regions, and some pixels that were
incorrectly tuned. No other cuts or masks are used inside the
region-of-interest. The resulting efficiencies as a function of the
global threshold are shown in figure \ref{fig:atlaspix1_efficiency}. The
efficiency stays almost constant and with the majority of the
measurements above \SI{99}{\percent} for the selected threshold range.
The error bars represent only the statistical uncertainty and do not
include systematic effects e.g. known-bad pixels and tuning effects.
Additional measurements indicate that the efficiency drops to
approximately \SI{70}{\percent} at a threshold of \SI{1.2}{\V}.

\section{Summary}

Multiple prototype sensors for the ATLAS ITk upgrade using ams H35 and
aH18 CMOS technology have been produced. The H35DEMO is a large scale
prototype that can be operated both as a monolithic system with an
integrated readout and as a hybrid sensor with a capacitively coupled
FE-I4 readout chip. The ATLASPix1 is a first large scale, monolithic
prototype designed specifically for the ATLAS ITk upgrade. Both
prototypes could be operated in a beam test setup with efficiencies
above \SI{99}{\percent}. This shows that large scale, CMOS pixel sensors
in ams technology can be build and operated with a variety of readout
options and are a suitable option for the outer layer of the ATLAS ITk
upgrade. The ATLASPix1 results only show its basic
functionality. Additional performance increases due to better tuning and
higher substrate resistivity are expected in the future. Systematic
measurements of irradiated ATLASPix1 prototypes with a variety of
irradiation sources are currently ongoing to verify previous
measurements of radiation hardness with small prototypes \cite{ccpdv4}.

\section*{Acknowledgments}

The authors gratefully acknowledge the support by the CERN PS and SPS
instrumentation team and Fermilab Test Beam Facilities. The authors
thank Andreas Nürnberg and Dominik Dannheim for fruitful discussions and
collaboration. The research presented in this paper was supported by the
SNSF grants 20FL20\_173601, 200021\_169015 and 200020\_169000.
This project has received funding from the European Union’s Horizon 2020
research and innovation programme under grant agreement No 675587.

\providecommand{\bibsection}{\section*{References}}
\bibliographystyle{elsarticle-num-names}
\bibliography{references.bib}

\end{document}